\makeatletter
\def\input@path{{}}
\makeatother
\documentclass[english,final]{aipproc}
\usepackage[T1]{fontenc}
\usepackage[latin1]{inputenc}
\usepackage{graphicx}

\makeatletter

\layoutstyle{6x9}

\usepackage{babel}
\makeatother
\begin{document}

\newcommand{\missET}{{E_{T}\!\!\!\!\!\!\!/\:\:\,}}

\classification{}

\author{Pavel M. Nadolsky}{address={High Energy Physics Division, Argonne National Laboratory, Argonne, IL 60439-4815, U.S.A.}}

\begin{flushright}Preprint ANL-HEP-CP-04-138, hep-ph/0412146 \end{flushright}

\title{Theory of $W$ and $Z$ boson production\thanks{Contribution to the proceedings of the 15th Topical Conference on Hadron Collider Physics (HCP 2004, June 14-18, 2004, East Lansing, MI)}}

\begin{abstract}
Success of precision studies in $W$ and $Z$ boson production at
the Tevatron and LHC depends on the progress in reducing theoretical
uncertainties to the level of $1-2\%$. I review recent developments
in theoretical understanding of $W$ and $Z$ boson production and
remaining issues to be solved to meet demands of the near-future experiments. 
\end{abstract}
\maketitle
Production of $W$ and $Z$ bosons at $p\bar{p}$ and $pp$ colliders
is well-suited for precision tests of hadronic matter and electroweak
interactions, due to its relatively simple mechanism and clean experimental
signatures. $W^{\pm}$ and $Z^{0}$ bosons will be mass-produced in
the Tevatron Run-2 and at the Large Hadron Collider (LHC) with the
goal to measure electroweak parameters, probe the internal structure
of nucleons, and monitor the collider luminosity. Deviations from
the predictions of the Standard Model may possibly occur in the $W$
and $Z$ boson data as a result of new gauge interactions, supersymmetry,
particle compositeness, or extra spatial dimensions. Searches for
new physics are not feasible without adequate evaluation of the large
irreducible background of $W$ and $Z$ bosons from the conventional
processes.

\textbf{The measurement of the $W$ boson mass $M_{W}$ and width
$\Gamma_{W}$} is an essential component of the Tevatron physics program.
In the Standard Model, radiative corrections relate $M_{W}$ to the
square of the top quark mass, $m_{t}^{2}$, and the \emph{logarithm}
of the Higgs boson mass, $\log(M_{H})$ \cite{Awramik:2003rn}.%
\footnote{$M_{W}$ also receives small corrections proportional to $M_{H}^{2}$.%
} A high-precision measurement of $M_{W}$ and $m_{t}$ constrains
the range of $M_{H}$ allowed within the Standard Model. For example,
if $M_{W}$ and $m_{t}$ are known within $~30$~MeV and 2~GeV,
respectively, $M_{H}$ is predicted within $35\%$. Conversely, comparison
of values of $M_{W},$ $m_{t},$ and $M_{H}$ measured in different
experiments and channels may reveal signs of new physics. The Run-2
plans to measure $M_{W}$ with accuracy $\approx30$ MeV per channel
per experiment, and $\Gamma_{W}$ with accuracy 25-30 MeV \cite{Brock:1999ep}
. The goal of LHC is to reduce $\delta M_{W}$ below 10-15 MeV in
order to compete with the precision measurement of $M_{W}$ at the
future Linear Collider \cite{Haywood:1999qg}.

At collider luminosities above $1\mbox{ fb}{}^{-1}$, theoretical
uncertainties (currently several percent of the cross section) become
the dominant source of error in $W$ and $Z$ observables. These uncertainties
must be reduced in order to measure the total cross sections with
accuracy $<2\%$ and $M_{W}$ with accuracy $<0.04\%$, as envisioned
by the Tevatron and LHC programs. Several dynamical factors contribute
at the percent level, including QCD radiative corrections of order
${\cal O}(\alpha_{s}^{2})$, electroweak corrections of order ${\cal O}(\alpha)$,
uncertainties in parton distributions, and power corrections to resummed
differential cross sections. Tangible progress has been made in exploring
these complex factors, mostly by considering each component separately
from the rest in order to simplify the problem. Correlations between
effects of different dynamical origins have been often neglected,
despite their probable significance in the full result. The percent-level
QCD and electroweak contributions will have to be integrated into
a future framework that would account for their correlations and be
practical enough to serve multifaceted purposes of $W$ and $Z$ physics.
In this talk, I review the recent advancements and future steps towards
development of such a comprehensive model.

\textbf{Overview of the process.} At hadron colliders, the massive
electroweak bosons are produced predominantly via $q\bar{q}$ annihilation
and detected by the decay into a pair of leptons. A typical $W$ or
$Z$ observable is affected by the QCD and electroweak radiation from
the hadronic initial state; production, propagation, and decay of
massive bosons; and electroweak radiation from the leptonic final
state. A hadronic cross section is evaluated in perturbative QCD as
a product of hard scattering cross sections and nonperturbative parton
distribution functions (PDF's). The hard scattering cross sections
are known at ${\cal O}(\alpha_{s}^{2})$ for inclusive observables
(depending on one energy scale), and at ${\cal O}(\alpha_{s})$ for
observables depending on several energy scales. Phenomenological parametrizations
of the PDF's can be taken from a global fit to hadronic data. Uncertainties
in the PDF's arising from diverse experimental and theoretical sources
can be propagated from the global analysis into the predictions for
the $W$ and $Z$ cross sections \cite{TungPDF}. When the soft QCD
radiation is enhanced in the scattering (\emph{e.g.}, when the transverse
momentum $q_{T}$ of the electroweak boson is much smaller than its
invariant mass $Q$), the finite-order perturbation theory has to
be replaced by resummation of large logarithmic corrections through
all orders of $\alpha_{s}$. The resummation is achieved by modifying
the form of QCD factorization in the affected regions in order to
absorb the problematic soft logarithms into exponential Sudakov form
factors.

The full ${\cal O}(\alpha)$ electroweak corrections are known both
for $W$ boson production \cite{Dittmaier:2001ay} and $Z$ boson
production \cite{Baur:2001ze}. They depend on the flavor of produced
leptons. Not all ${\cal O}(\alpha)$ corrections are of equal importance
throughout all phase space. The resonant production of $W$ bosons
at $Q\approx M_{W}$ is described well within {}``the pole approximation''
\cite{Baur:1998kt}, which neglects the subleading $WZ$ box diagrams.
For its part, the pole approximation is dominated by virtual corrections
to boson propagators and vertices, as well as by QED radiation off
the final-state charged leptons \cite{Berends:1984qa}. The QED radiation
from the initial-state quarks and interference between the initial
and final states are of a smaller magnitude numerically, but may have
some impact on precision measurements. Away from the $W$ boson resonance
($Q\gg M_{W}$), the pole approximation becomes unreliable and has
to be replaced by the full ${\cal O}(\alpha)$ result \cite{Baur:2004ig}.
Yet higher-order electroweak corrections may be non-negligible as
well, especially if enhanced by large logarithms near the boundaries
of available phase space or at large boson virtualities. $W$ and
$Z$ boson production with radiation of two additional photons has
been studied in Ref.~\cite{Baur:1999hm}. Leading contributions from
radiation of multiple soft photons can be resummed through all orders
of $\alpha$, as it was done recently for $W$ decays in Refs.~\cite{Placzek:2003zg,CarloniCalame:2004qw}.

\begin{figure}
\includegraphics[%
  width=0.48\columnwidth,
  height=6cm,
  keepaspectratio]{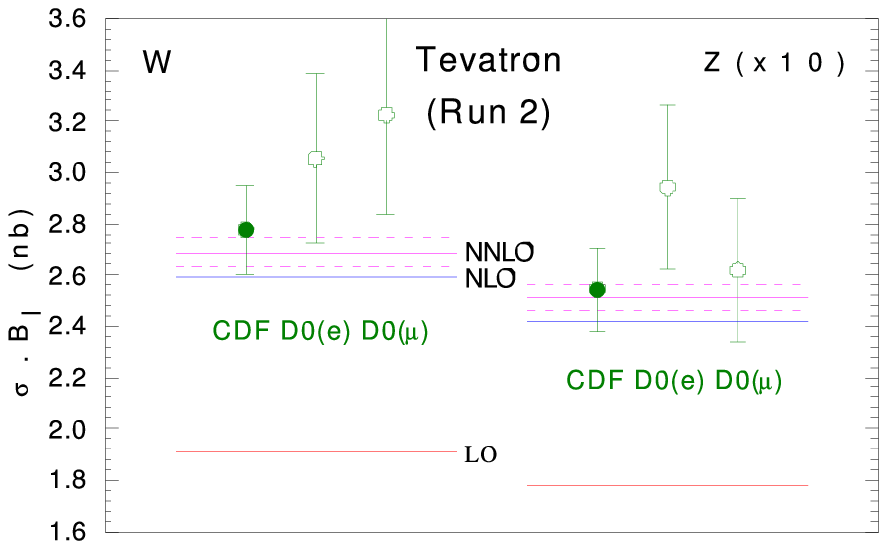}~\includegraphics[%
  width=0.48\columnwidth]{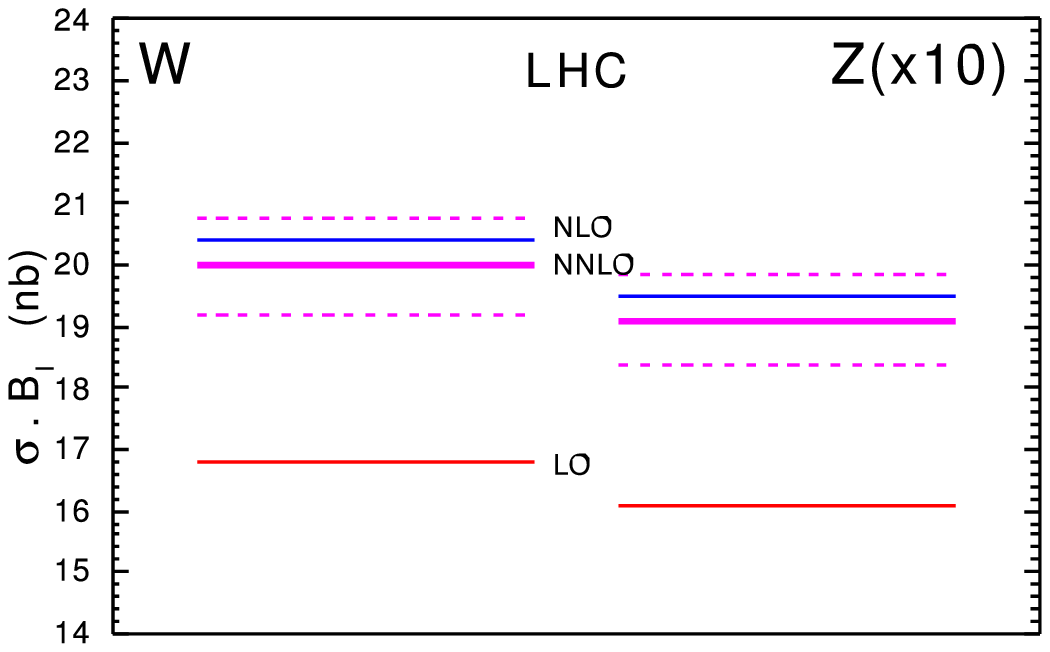}

\caption{\label{fig:sigtot1} Effect of the ${\cal O}(\alpha_{s}^{2}$) (NNLO)
corrections on the total cross sections $\sigma_{tot}(pp\hspace{-6pt}{}^{{}^{(-)}}\rightarrow(W\rightarrow\ell\nu)X)$
and $\sigma_{tot}(pp\hspace{-6pt}{}^{{}^{(-)}}\rightarrow(Z\rightarrow\ell\bar{\ell})X)$
at the Tevatron and LHC \cite{Martin:2003sk}. The narrow-width EW
approximation was used.}
\end{figure}

\textbf{Total cross sections}. The interplay of various factors can
be most easily demonstrated on the example of the total cross sections,
given in perturbative QCD by\begin{eqnarray}
\sigma_{tot}(pp\hspace{-8pt}{}^{{}^{(-)}}\rightarrow(V\rightarrow\ell_{1}\ell_{2})X) & = & \sum_{a,b=q,\bar{q},g}\int dx_{1}dx_{2}f_{a/p}(x_{1},\mu)f_{b/p\hspace{-6pt}{}^{{}^{(-)}}}(x_{2},\mu)\nonumber \\
 & \times & \widehat{\sigma}_{tot}(ab\rightarrow(V\rightarrow\ell_{1}\ell_{2})X).\end{eqnarray}
Here $V=W$ or $Z$; $\ell_{1}\ell_{2}$ are the charged lepton and
neutrino ($\ell\nu$) in $W$ boson production, or the charged lepton
and antilepton ($\ell\bar{\ell}$) in $Z$ boson production; $\widehat{\sigma}_{tot}$
are the hard scattering cross sections, known to ${\cal O}(\alpha_{s}^{2})$
\textit{}\cite{Hamberg:1990np,Harlander:2002wh}\textit{;} and \textit{}$f_{a/p}(x,\mu)$
are the parton distributions. The integration is over the partonic
momentum fractions $x_{1}$ and $x_{2}$, and the summation is over
the relevant parton flavors. The $W$ and $Z$ total cross sections
have been evaluated recently by the MRST group by using the parton
distributions from a recent global analysis, realized with inclusion
of all known next-to-next-to-leading-order, or NNLO, corrections \cite{Martin:2003sk}.
The ${\cal O}(\alpha_{s}^{2})$ correction was found to increase $\sigma_{tot}$
by about $4\%$ at the Tevatron and reduce $\sigma_{tot}$ by about
$2\%$ at the LHC {[}cf.~Fig.~\ref{fig:sigtot1}{]}. The dependence
of $\sigma_{tot}$ on the arbitrary factorization scale $\mu$ is
reduced at ${\cal O}(\alpha_{s}^{2})$ to about $1\%$, suggesting
high stability with respect to QCD corrections of order $\alpha_{s}^{3}$
and beyond. 

The magnitude of ${\cal O}(\alpha)$ electroweak corrections is comparable
to that of the QCD corrections. In the limit of the vanishing boson's
width $\Gamma_{V}$, the production of the vector bosons can be separated
from their decay, by introducing on-shell production cross sections
$\sigma_{tot}(pp\hspace{-8pt}{}^{{}^{(-)}}\rightarrow V)$ and branching
ratios $\mbox{{Br}}(V\rightarrow\ell_{1}\ell_{2})$: \begin{equation}
\sigma_{tot}(pp\hspace{-8pt}{}^{{}^{(-)}}\rightarrow(V\rightarrow\ell_{1}\ell_{2})X)=\sigma_{tot}(pp\hspace{-8pt}{}^{{}^{(-)}}\rightarrow V)\cdot\mbox{{Br}}(V\rightarrow\ell_{1}\ell_{2}).\label{Branch}\end{equation}
At this level of accuracy, the branching ratio defined via Eq.~(\ref{Branch})
coincides with the partial width, $\mbox{{Br}}(V\rightarrow\ell_{1}\ell_{2})=\Gamma(V\rightarrow\ell_{1}\ell_{2})/\Gamma_{V}$.
The partial width can be defined beyond the Born level as a universal
pseudo-observable, included in concrete cross sections via process-specific
relations. For instance, the partial widths for $Z$ bosons are incorporated
in the fit to the LEP $Z$-pole observables with the help of a complicated
procedure, which accounts for the $Z$ line shape, initial- and final-state
electroweak radiation, and $\gamma Z$ interference (see {}``The
$Z$ boson'' review in Ref.~\cite{Eidelman:2004wy}). An analogous
(but not identical) procedure relates the partial widths to the Tevatron
cross sections. A tree-level estimate in the Standard Model yields
$\Gamma(W\rightarrow\ell\nu)/\Gamma_{W}=11\%$ and $\Gamma(Z\rightarrow\ell\bar{\ell})/\Gamma_{Z}=3.3\%$,
while the Review of Particle Physics quotes $(10.68\pm0.12)\%$ and
$(3.3658\pm0.0023)\%$ based on the fit to the world (predominantly
LEP) data \cite{Eidelman:2004wy}. The shown RPP values are averaged
over three lepton generations. Consequently the narrow-width approximation
may deviate from the true Tevatron cross sections by up to $\sim3\%$,
if the Tevatron-specific ${\cal O}(\alpha)$ corrections are not taken
into account.

In addition, the narrow-width approximation neglects the Breit-Wigner
line shape of vector bosons and is not suitable for computation of
differential distributions. The next level of refinement is realized
in the effective Born approximation (EBA), which includes the final
width and radiative corrections to the electroweak couplings and boson's
propagator (but no effects of real particle radiation). With (without)
lepton identification requirements, the EBA $W$ cross section in
Run-2 exceeds the full ${\cal O}(\alpha)$ cross section by $2\%$
($4\%$) in the $e\nu_{e}$ decay channel, and $6\%$ ($2\%$) in
the $\mu\nu_{\mu}$ decay channel \cite{Baur:2004ig}. As can be seen
from this discussion, the electroweak corrections may induce sizable
variations in $\sigma_{tot}$. A percent-level comparison of the total
cross sections is only meaningful if the employed electroweak corrections
are well-documented and consistent. 

\begin{figure}
\includegraphics[%
  width=0.45\columnwidth,
  keepaspectratio]{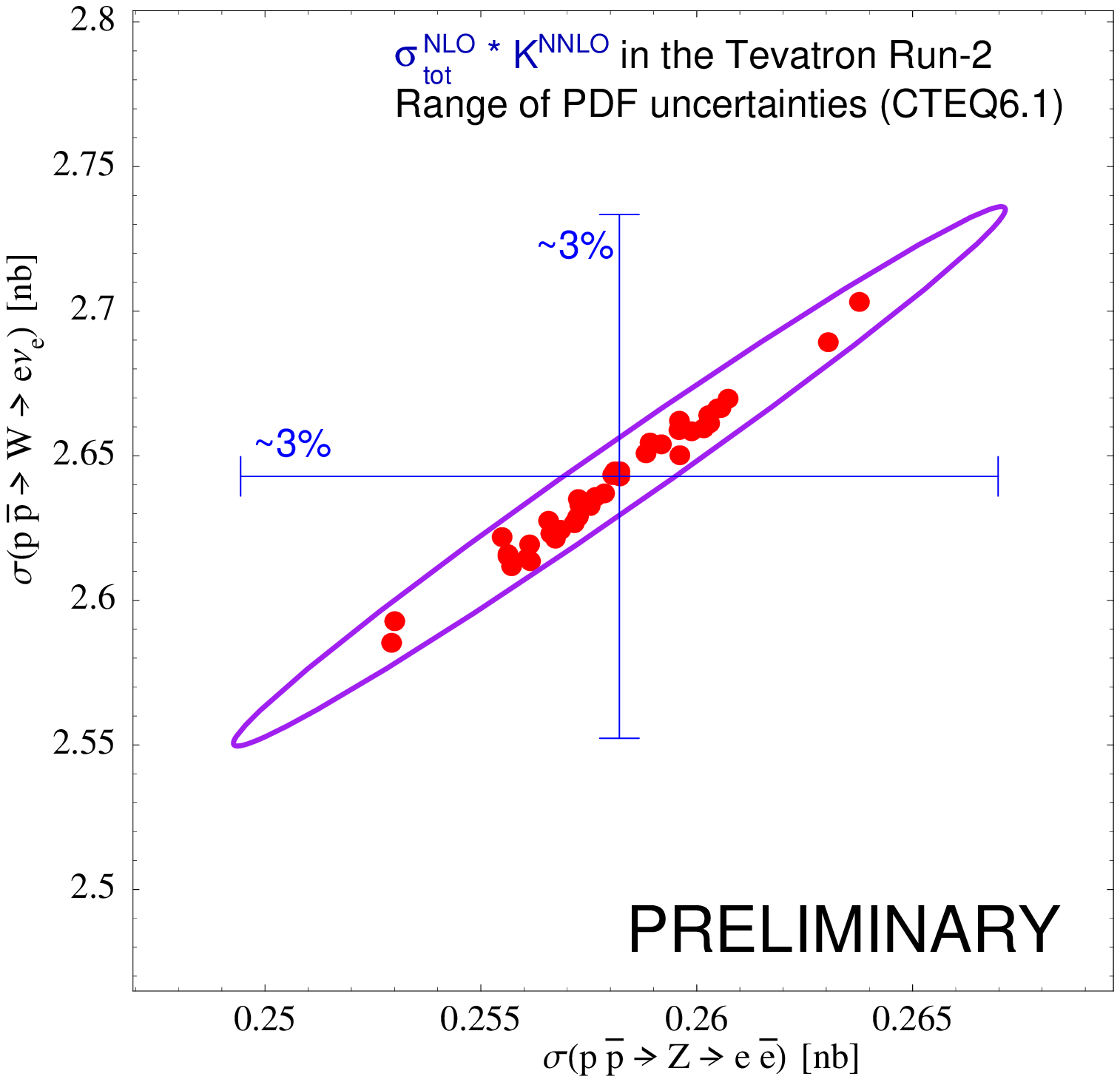}\includegraphics[%
  width=0.45\columnwidth,
  keepaspectratio]{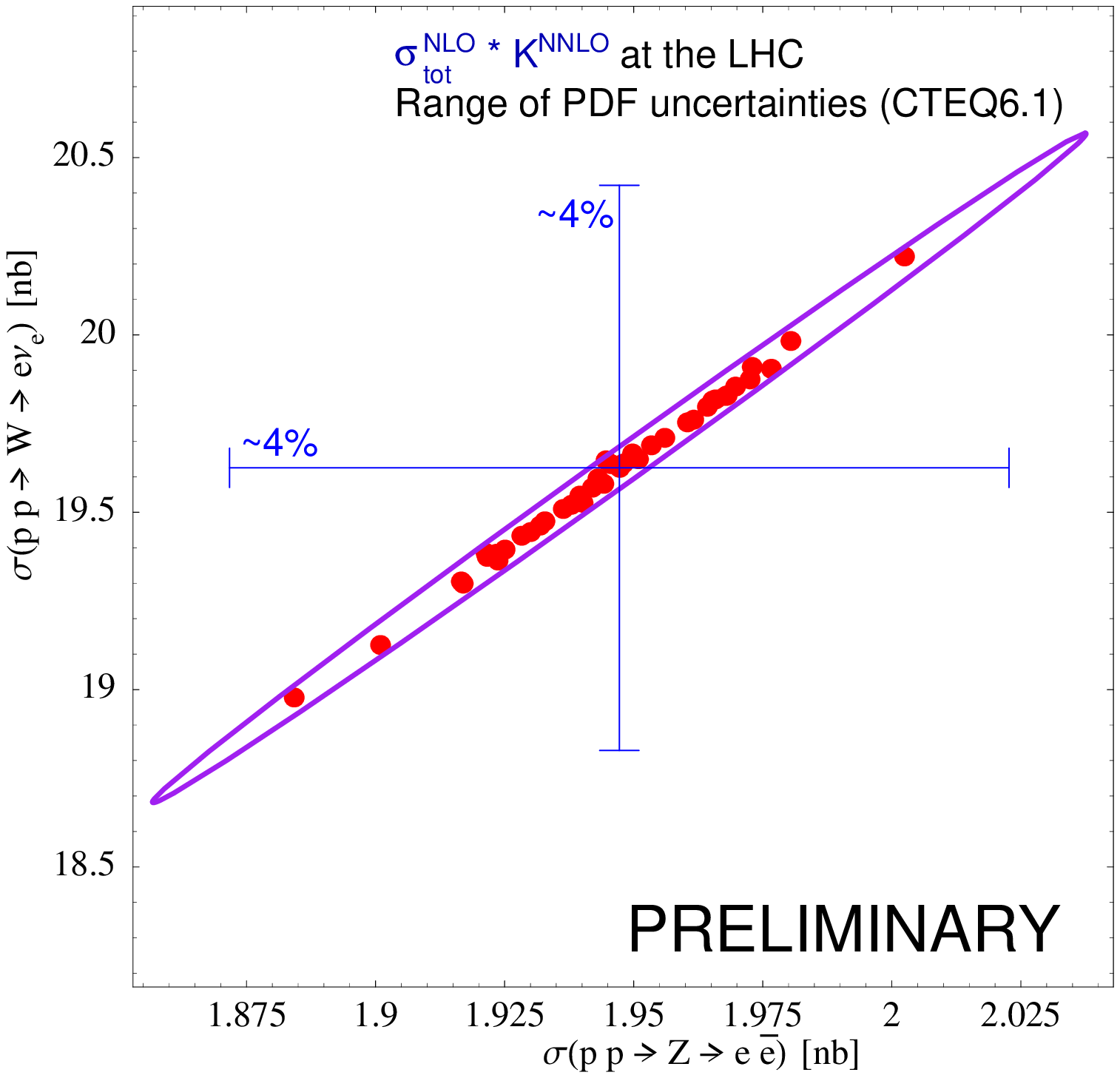}\vspace{-13pt}

\caption{\label{fig:WZCTEQ} Correlations between the total $W$ and $Z$
cross sections at the Tevatron and LHC found in the new CTEQ analysis
\cite{HustonSigTot}. The cross sections are computed at ${\cal O}(\alpha_{s})$
and rescaled by the NNLO $K$-factor, extracted from Ref.~\cite{Martin:2003sk}.
The dots correspond to 41 PDF sets from the CTEQ6.1 analysis \cite{Stump:2003yu}. }
\end{figure}

Uncertainty in $\sigma_{tot}$ due to imprecise knowledge of parton
distributions can be evaluated using one of the available methods
\cite{TungPDF}. The exact values of the PDF errors depend on the
conventions adapted by each group, but all lie in a few-percent range.
The CTEQ estimate (based on the most conservative criterion among
all groups) quotes the PDF error of $3\%$ ($4\%$) at the Tevatron
(LHC). The PDF errors for $W$ and $Z$ total cross sections are highly
correlated {[}cf.~Fig.~\ref{fig:sigtot1}{]}, reflecting the striking
similarity between the quark-dominated initial states of the two processes.
The ratio of the two cross sections, $R_{W/Z}=\sigma_{tot}(W)/\sigma_{tot}(Z)$,
is essentially invariant with respect to the changes in the parton
distributions. A measurement of $Z$ boson cross section can tightly
constrain the PDF input in $W$ boson production and other kindred
processes. Furthermore, fluctuations in the $Z$ boson rate in the
course of the collider run would reflect changes in the collider luminosity,
which can be controlled through the measurement of $\sigma_{tot}(Z)$
to the level of $1-2\%$. These nice features of the $W$ and $Z$
boson total cross sections advance them as likely {}``standard candle''
monitors of the partonic and collider luminosities at the LHC \cite{Dittmar:1997md}.
Of course, all percent-level effects must be put together in one calculation
to realize this goal. To reconstruct the total cross section from
the visible events, an accurate model of the detector acceptance is
needed. The $W$ boson acceptances can be controlled at the level
of $2\%$ by incorporating the parton-level NLO cross sections into
parton showering programs \cite{Frixione:2004us}.

\begin{figure}
\includegraphics[%
  width=0.48\columnwidth,
  keepaspectratio]{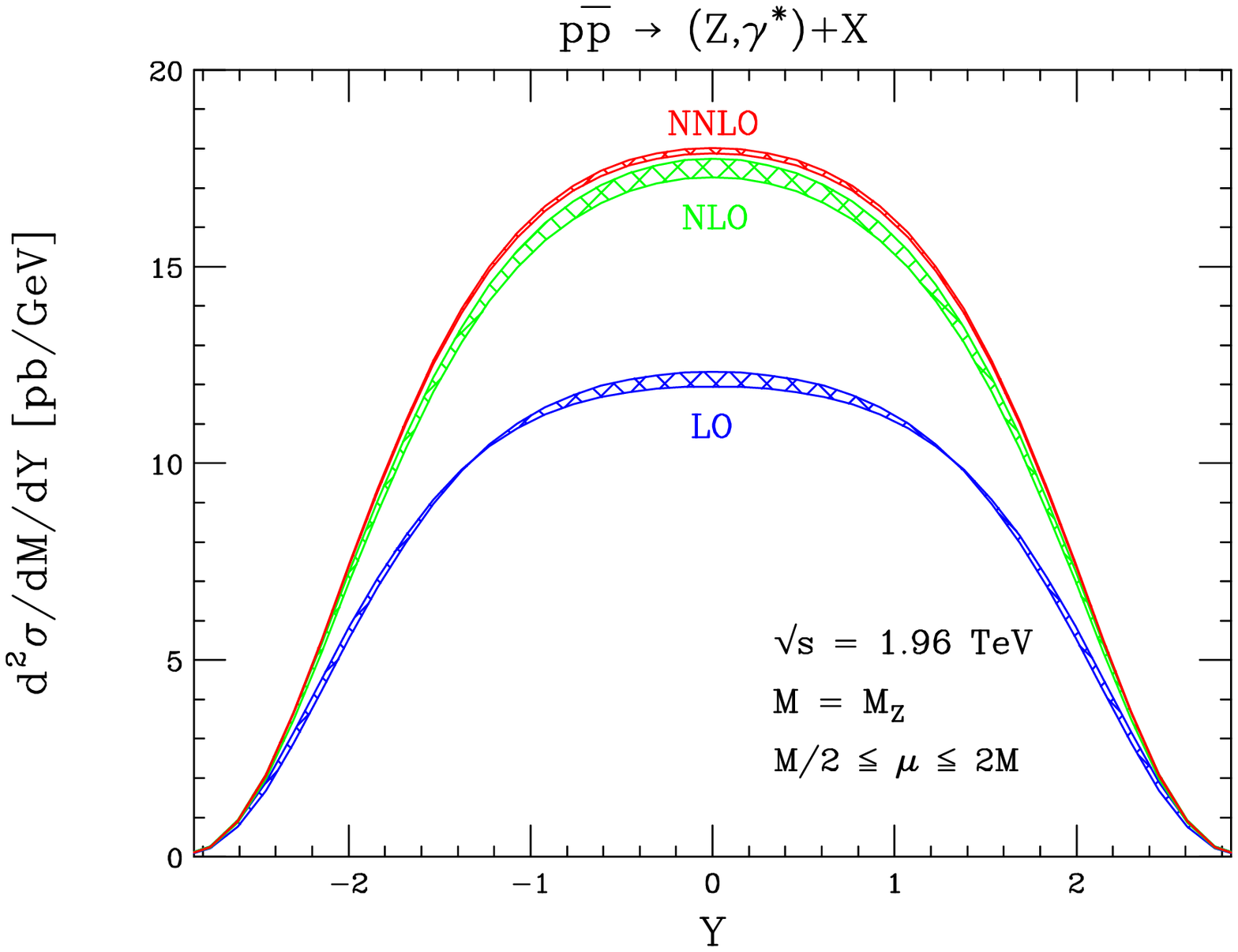}~~\includegraphics[%
  width=0.48\columnwidth,
  keepaspectratio]{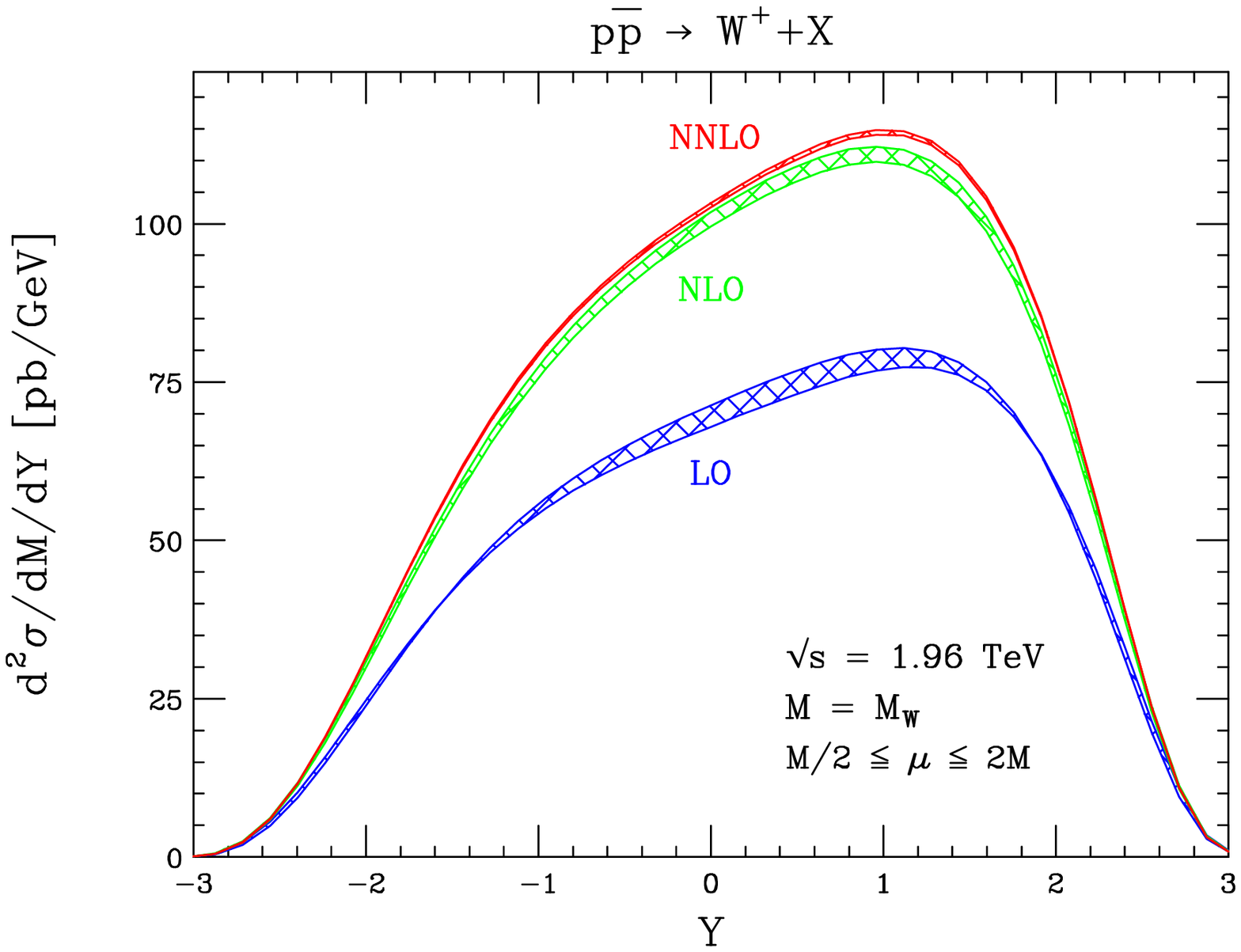}

\caption{\label{fig:RapRun2}${\cal O}(\alpha_{s}^{2})$ rapidity distributions
in the Tevatron Run-2 \cite{Anastasiou:2003ds}.}
\end{figure}

\textbf{Rapidity distributions} of massive electroweak bosons have
been recently computed at ${\cal O}(\alpha_{s}^{2})$ with the help
of a novel technique based on the optical theorem and recursive reduction
of arbitrary loop integrals to a small set of known {}``master integrals''
\cite{Anastasiou:2003ds}\textit{.} In a large range of rapidities
($\left|y\right|<2$), the ${\cal O}(\alpha_{s}^{2})$ correction
leads to essentially uniform enhancement of the cross section by $3-5\%$
in $Z$ production and $2.5-4\%$ in $W$ production {[}cf.~Fig.~\ref{fig:RapRun2}{]}.
The ${\cal O}(\alpha_{s}^{2})$ corrections are augmented and less
uniform in the forward rapidity regions. Similarly to the total cross
sections, the scale dependence is reduced in ${\cal O}(\alpha_{s}^{2})$
rapidity distributions below $1\%$.

\begin{figure}
\includegraphics[%
  width=0.49\columnwidth,
  keepaspectratio]{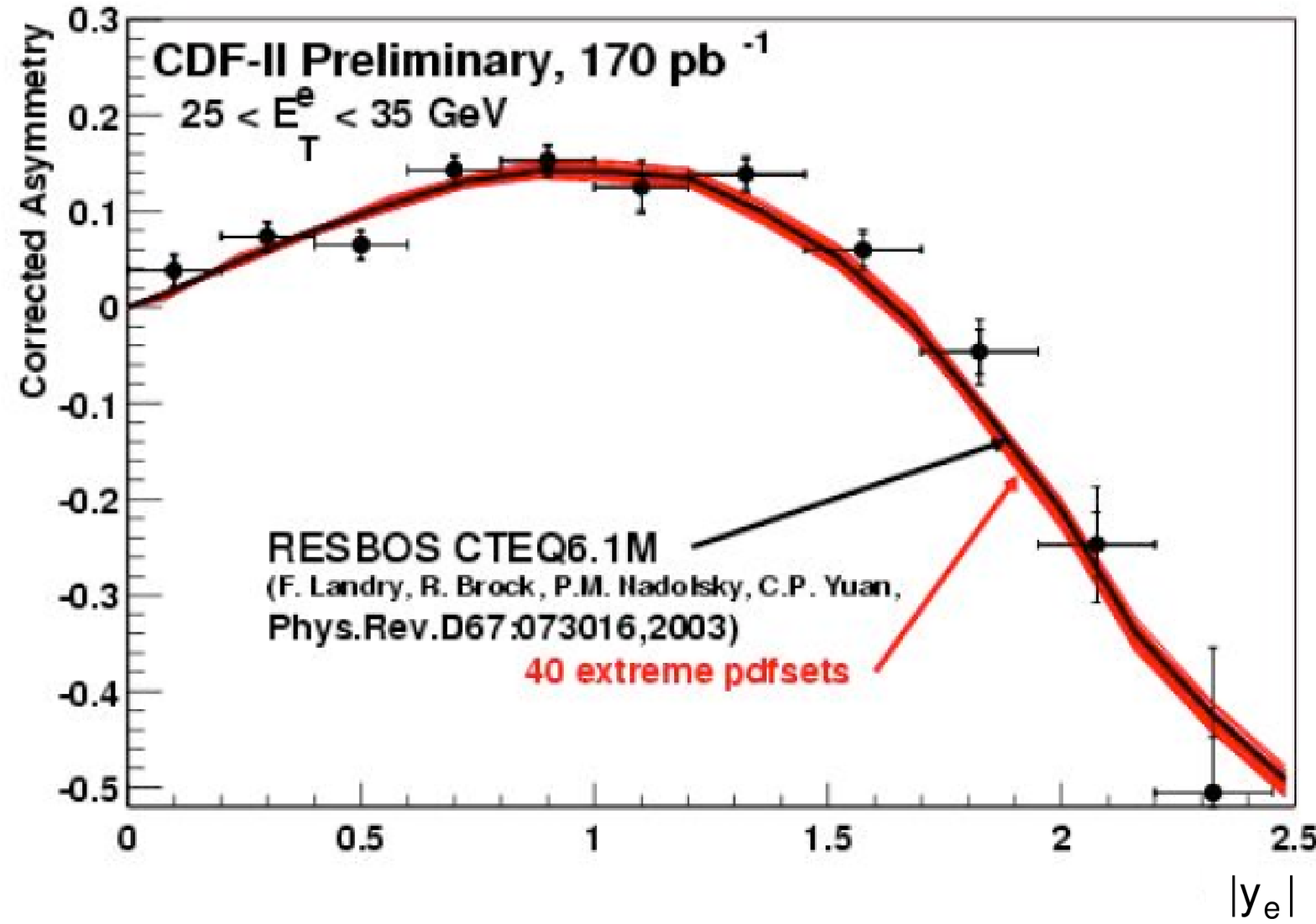}\includegraphics[%
  width=0.49\columnwidth,
  keepaspectratio]{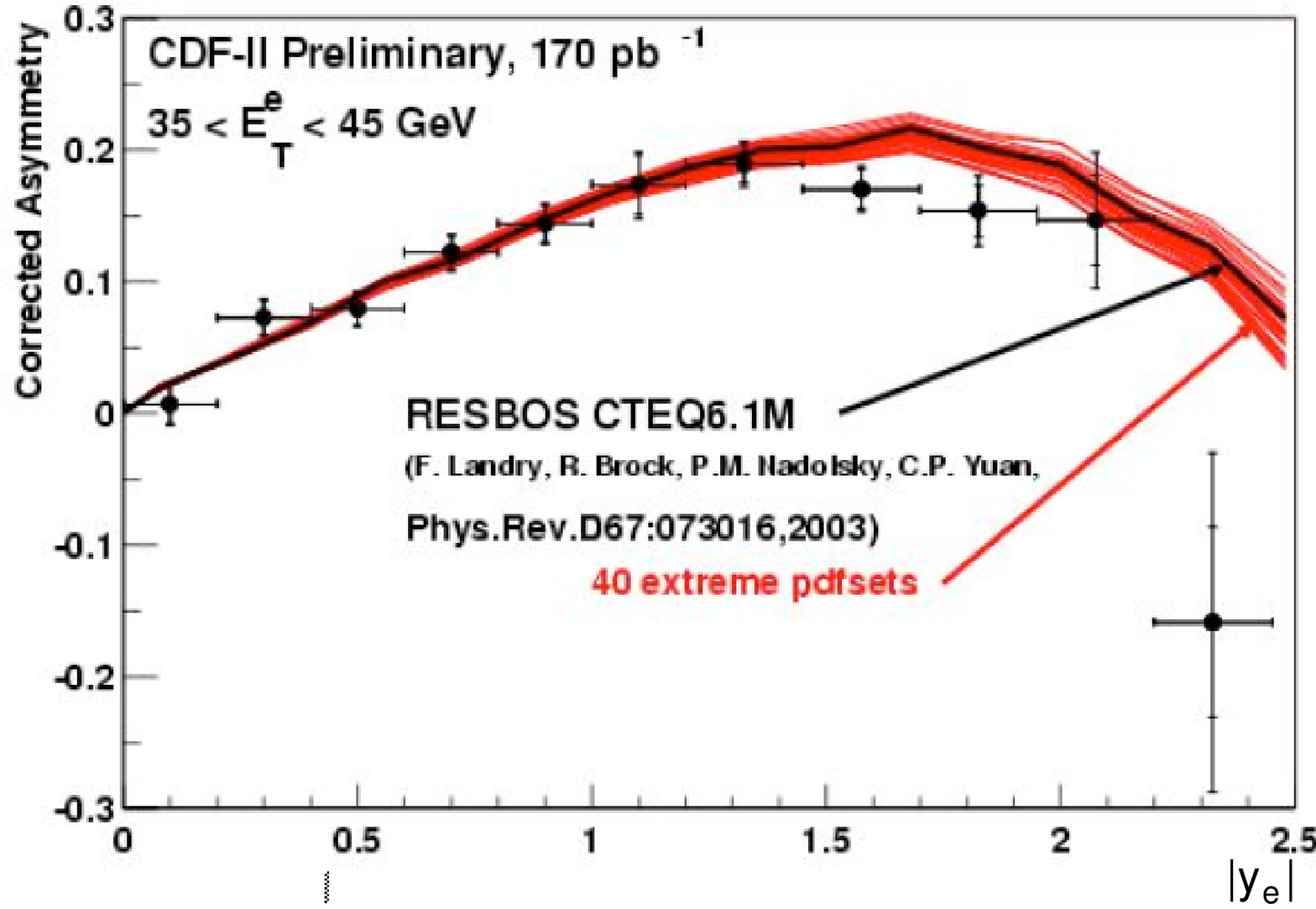}

\caption{\label{fig:Run2charge}The CDF Run-2 charge asymmetry vs. the resummed
result from the $q_{T}$ resummation program \textsc{ResBos} \cite{Balazs:1997xd,Landry:2002ix}.
The bands reflect the PDF uncertainty, evaluated using the CTEQ6.1
PDF set \cite{Stump:2003yu}.}
\end{figure}

The shape of the rapidity distributions is determined by the $x$
dependence of parton densities. $W$ boson production is sensitive
to the flavor composition of the PDF's because of the mixing of quark
flavors in $Wq\bar{q}$ coupling. At the $p\bar{p}$ collider Tevatron,
$W$ bosons predominantly probe the distributions of $u$ and $d$
quarks (recall that the distribution of $d$ antiquarks in an antiproton
is equal to the distribution of $d$ quarks in a proton). The CTEQ
and MRST analyses include the Tevatron $W$ cross sections in the
form of the charge asymmetry \[
A_{ch}(y_{\ell})\equiv\frac{d\sigma^{W^{+}}/dy_{\ell}-d\sigma^{W^{-}}/dy_{\ell}}{d\sigma^{W^{+}}/dy_{\ell}+d\sigma^{W^{-}}/dy_{\ell}},\]
 defined in terms of the rapidity distributions $d\sigma^{W^{\pm}}/dy_{\ell}$
of high-$p_{T}$ charged leptons from $W^{\pm}$ boson decays \cite{Berger:1988tu}.
At forward $y_{\ell}$, the Tevatron charge asymmetry probes the ratio
$d(x,M_{W})/u(x,M_{W})$ at $x$ up to about $0.3$. In the experiment,
selection cuts are imposed on the transverse energies $E_{T}$ of
the charged leptons and neutrinos as a means to suppress the backgrounds
and discriminate between different ranges of $x$. The preliminary
CDF Run-2 asymmetry is in a good agreement with the resummed theory
cross section at $25<E_{T}^{e}<35$ GeV {[}Fig.~\ref{fig:Run2charge}(a){]},
but some points fall out of the CTEQ PDF uncertainty band at $35<E_{T}^{e}<45$
GeV {[}Fig.~\ref{fig:Run2charge}(b){]}, indicating the need in further
improvements of the PDF model.

\begin{figure}
\includegraphics[%
  height=6cm,
  keepaspectratio]{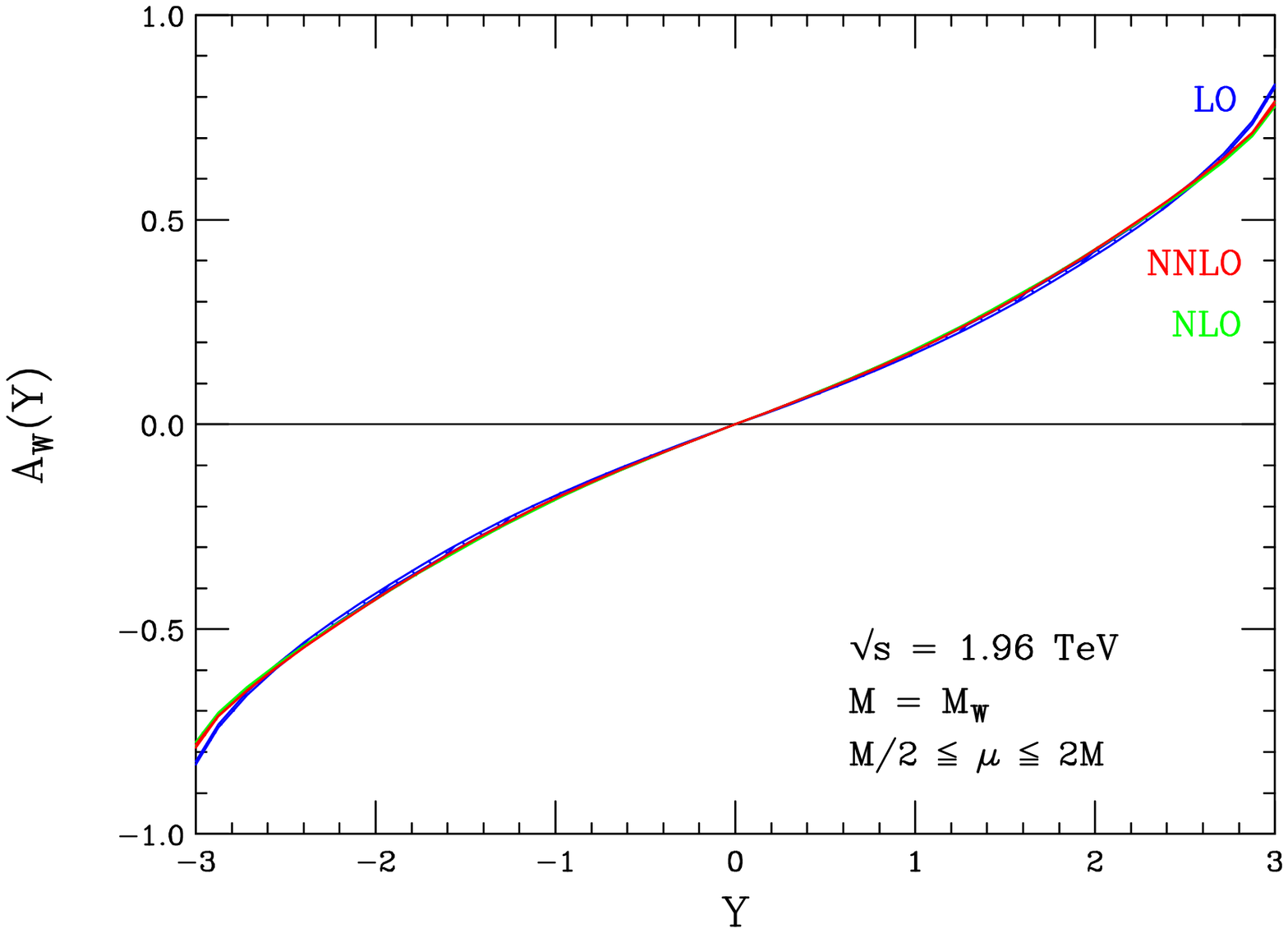}~\includegraphics[%
  height=6cm,
  keepaspectratio]{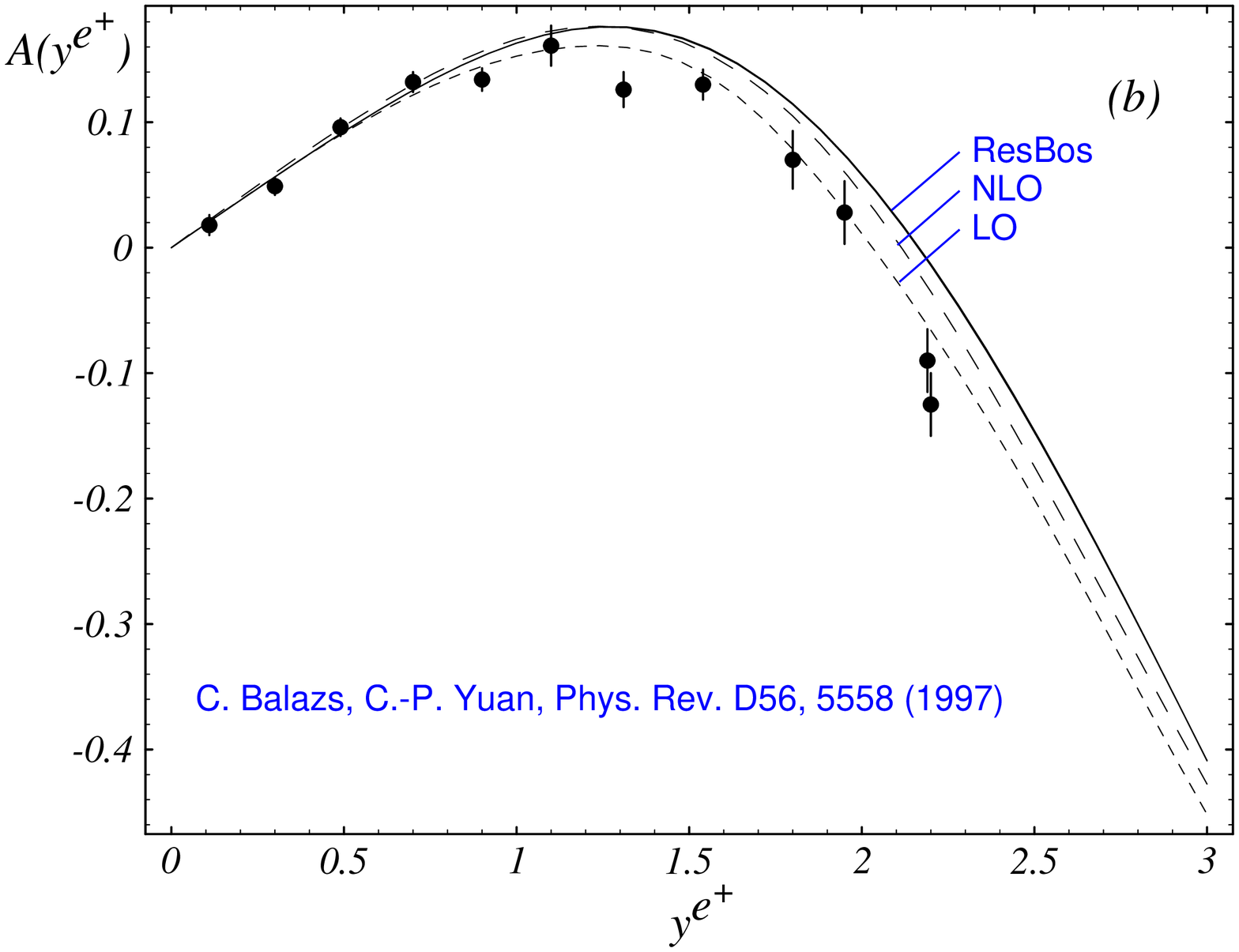}

\caption{(a) Charge asymmetry $A_{ch}(y)$ of $W$ boson rapidity distributions,
$d\sigma/dy$, at various orders of $\alpha_{s}$ (no acceptance cuts
are imposed) \cite{Anastasiou:2003ds}; charge asymmetry $A_{ch}(y_{\ell})$
in the Tevatron Run-1 for $q_{T}<30$ GeV, $p_{T}^{e^{+},\nu}>25$
GeV in the leading-order, next-to-leading order, and resummed calculations
\cite{Balazs:1997xd}.\label{fig:AchCuts}}
\end{figure}

If the transverse momentum $q_{T}$ of $W$ boson is neglected, a
simple kinematic relationship exists between the rapidity $y_{\ell}$
and transverse momentum $p_{T\ell}$ of the charged lepton. This relationship
induces a strong correlation between the shapes of $A_{ch}(y_{\ell})$
and $d\sigma/dp_{T\ell}$, so that a measurement of $A_{ch}(y_{\ell})$
reduces the PDF uncertainty in $d\sigma/dp_{T\ell}$ and the measured
$W$ boson mass \cite{Stirling:1989vx}. While being a helpful approximation,
this correlation does not hold exactly, due to non-zero $q_{T}$ of
$W$ bosons and acceptance constraints on the leptonic momenta. If
no acceptance cuts are imposed, the charge asymmetry is essentially
invariant with respect to the QCD corrections {[}Fig.~\ref{fig:AchCuts}(a){]}.
The acceptance cuts re-introduce the sensitivity of $A_{ch}(y_{\ell})$
to QCD corrections, due to the differences in phase space available
at different orders of perturbative calculation. In Fig.~\ref{fig:AchCuts}(b),
differences can be noticed between the leading-order, next-to-leading-order,
and resummed cross sections, caused by different shapes of $q_{T}$
distributions in the three calculations. The differences are augmented
at forward rapidities, where the convergence of perturbation series
is reduced by enhancement of near-threshold radiation. These features
suggest that the ${\cal O}(\alpha_{s}^{2})$ corrections cannot be
replaced by a constant $K$-factor at forward rapidities. An additional
study may be needed to evaluate the impact of QCD radiation and phase
space restrictions on the observed $A_{ch}(y_{\ell})$. 

\textbf{$p_{T}$ distributions and extraction of $M_{W}$.} The line
shape of $W$ bosons is not observed in the leptonic channels because
of the missing information about the longitudinal momentum of the
neutrino. For this reason, the $W$ boson mass is commonly deduced
from the distribution of the transverse momenta of the decay leptons.
The distribution in the leptonic transverse mass, $M_{T}^{\ell\nu}\equiv\sqrt{\left(\left|\vec{p}_{T\ell}\right|+\left|\vec{p}_{T\nu}\right|\right)^{2}-\left(\vec{p}_{T\ell}+\vec{p}_{T\nu}\right)^{2}}$,
was the preferred observable to extract $M_{W}$ in Run-1 because
of its reduced sensitivity to the mechanism of $W$ boson production.
The kinematical (Jacobian) peak in the $M_{T}^{\ell\nu}$ distribution
is located exactly at $M_{T}^{\ell\nu}=M_{W}$ at the Born level and
smeared by the $W$ boson's width and radiative corrections. The shape
of the Jacobian peak is not affected, to the first order, by the transverse
motion of $W$ bosons, caused predominantly by their recoil against
the soft QCD radiation. For this reason, extraction of $M_{W}$ from
a fit to the $M_{T}^{\ell\nu}$ distribution is less reliant on precise
knowledge of the $q_{T}$ spectrum of $W$ bosons and PDF's, which
is crucial in the other methods \cite{Smith:1983aa}. However, the
accuracy of the $M_{T}^{\ell\nu}$ method suffers from the limited
precision in the measurement of the neutrino's transverse momentum
(equated in the experiment to the missing transverse energy, $E_{T}\hspace{-13pt}/\hspace{8pt}$).
An alternative technique extracts $M_{W}$ from the transverse momentum
distribution of the charged lepton, $d\sigma/dp_{T\ell}$. This method
does not suffer from the complications associated with the reconstruction
of $E_{T}\hspace{-13pt}/\hspace{8pt}$, but, as a trade-off, it requires
to precisely know the $q_{T}$ distribution. The direct measurement
of $d\sigma/dq_{T}$ involves reconstruction of $E_{T}\hspace{-13pt}/\hspace{8pt}$
and has low precision. A better prediction for the $q_{T}$ spectrum
can be made within the theoretical model by applying the resummation
methods described below. The $W$ boson mass can be also extracted
from the $p_{T}$ distribution of the neutrinos, or the ratios $[d\sigma/M_{T}^{\ell\nu}(W)]/[d\sigma/M_{T}^{\ell\bar{\ell}}(Z)]$
and $\sigma_{tot}(W)/\sigma_{tot}(Z)$ of $W$ and $Z$ transverse
mass distributions and total cross sections \textit{}\cite{Brock:1999ep,Rajagopalan:1996wx,Giele:1998uh,Shpakov:2000eq}.
The systematical error tends to be smaller in the last two methods
because of the cancellation of common uncertainties in the ratio of
$W$ and $Z$ cross sections. However, the reduced systematical uncertainty
is balanced in the full result by a larger statistical error, propagated
into $M_{W}$ from a smaller cross section for $Z$ boson production.
The uncertainties on $M_{W}$ will be comparable in all methods towards
the end of Run-2, and several techniques will be probably employed
by the experimental collaborations as a way to cross check the systematics.

\textbf{Factorization at small transverse momenta.} The reliability
of the finite-order cross section is jeopardized at small $q_{T}$
by incomplete cancellation of soft QCD singularities, which leaves
large logarithms $\ln^{n}(q_{T}/Q)$ in the hard matrix elements.
A re-arrangement of the perturbation series cures the instability
of the theory at $q_{T}^{2}\ll Q^{2}$ by summing the troublesome
$q_{T}$ logarithms through all orders of $\alpha_{s}$ into a soft
(Sudakov) form factor \cite{Dokshitzer:1978yd}. The validity of such
re-arrangement is proved by a factorization theorem in the method
by Collins, Soper, and Sterman (CSS) \cite{Collins:1985kg}. The resummation
in vector boson production is a special case of a more general problem,
and essentially the same method applies to hadroproduction in $e^{+}e^{-}$
scattering \cite{Collins:1981uk}, and semi-inclusive hadroproduction
in deep-inelastic scattering \cite{Collins:1993kk,Meng:1996yn,Nadolsky:1999kb}.
The CSS formalism automatically preserves the fundamental symmetries
(renormalization- and gauge-group invariance, energy-momentum conservation)
and is convenient in practice. The $q_{T}$ resummation can be extended
to include effects of particle thresholds \cite{Kulesza:2002rh},
heavy quark masses \cite{Nadolsky:2002jr}, and hadronic spin \cite{Weber:1991wd,Nadolsky:2003fz}.
\textsc{ResBos} \cite{Balazs:1997xd,Landry:2002ix} is a Monte-Carlo
integrator program that quickly and accurately evaluates the CSS resummed
cross sections in Drell-Yan-like processes.

\begin{figure}
\includegraphics[%
  height=7cm,
  keepaspectratio]{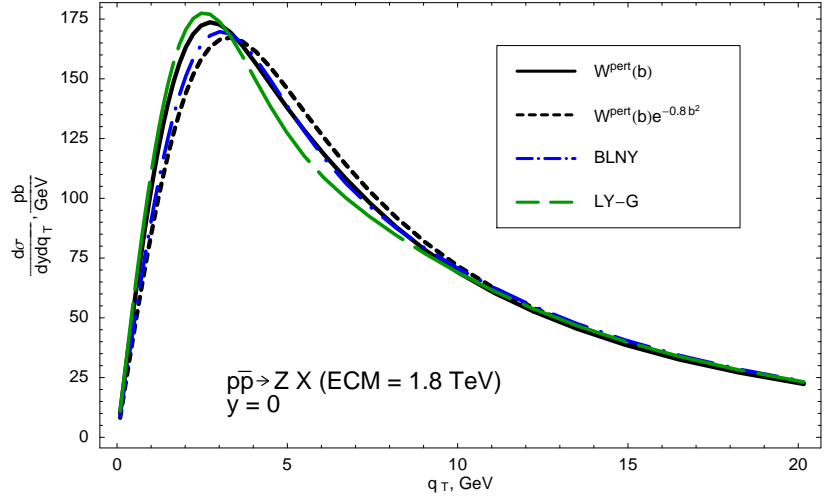}

\caption{\label{fig:ResZ} The CSS resummed cross sections in $Z$ boson production
at the Tevatron. The curves are computed in several models for the
CSS form factor $W(b)$ at large impact parameters ($b>1\mbox{{ GeV}}{}^{-1}$):
(a)~$W(b)$ at large $b$ is given by extrapolation of its perturbative
part from $b<1\mbox{{ GeV}}{}^{-1}$ (solid); (b)~the same as (a),
multiplied by a Gaussian smearing term $e^{-0.8b^{2}}$(short-dashed);
(c) a phenomenological BLNY form, which shows good agreement with
the Run-1 $Z$ data (dot-dashed) \cite{Landry:2002ix}; (d) an updated
Ladinsky-Yuan form, which shows worse agreement with the Run-1 $Z$
data (long-dashed) \cite{Landry:2002ix}. Note that the extrapolation
model (curves (a) and (b)) must include a Gaussian smearing term $e^{-gb^{2}},$
with $g\sim0.8\mbox{{ GeV}}^{2}$, in order to be close to the BLNY
form (and, hence, to the data). }
\end{figure}

All small-$q_{T}$ logarithms arise in the CSS method from the form
factor $W(b)$ in impact parameter ($b)$ space, composed of the Sudakov
exponential and $b$-dependent parton distribution functions. The
resummed $q_{T}$ distribution is obtained by taking the Fourier-Bessel
transform of $W(b)$ into $q_{T}$ space (realized numerically in
\textsc{ResBos}). The alternative approaches evaluate the Fourier-Bessel
transform of the leading logarithmic towers analytically, with the
goal to improve transition from the resummed cross section to the
finite-order cross section at intermediate $q_{T}$ \cite{Ellis:1998ii,Kulesza:2001jc}.
 The integration over all $b$ in the Fourier-Bessel transform introduces
sensitivity to the nonperturbative QCD dynamics at $b>1\mbox{{ GeV}}^{-1}$.
In $W$ and $Z$ boson production, the nonperturbative terms are strongly
suppressed by the shape of $W(b)$, as well as by the oscillations
in the Fourier-Bessel transform integrand at large $q_{T}$. Nonetheless,
mild sensitivity to the large-$b$ behavior remains at $q_{T}<10$
GeV, contributing on the top of the overall shape that is tightly
constrained by the perturbative contributions. The position of the
peak in $d\sigma/dq_{T}$ can move by several hundred MeV depending
on the choice of the nonperturbative model {[}cf.~Fig.~\ref{fig:ResZ}{]}.
The range of uncertainty in the $W$ boson's $q_{T}$ distribution
is substantially reduced by requiring the nonperturbative model to
agree with the measured $q_{T}$ distribution in $Z$ boson production.
In Run-1, the uncertainty in $M_{W}$ due to incomplete knowledge
of the $q_{T}$spectrum of $W$ bosons was estimated to amount to
15-20 MeV.

While significant effort has been put into the study of $W(b)$ at
large $b$ \cite{Kulesza:2002rh,Qiu:2000hf,Guffanti:2000ep,Ji:2004xq},
none of the existing approaches was able so far to adequately describe
the observed $Z$ boson distribution without introducing free parameters.%
\footnote{It is worth pointing out that the global features of $W$ and $Z$
$q_{T}$ distributions are uniquely determined by the perturbative
contributions and agree well in the different approaches. The challenging
part is to describe mild variations at $q_{T}$ below 10 GeV, which
cannot be neglected in high-precision measurements.%
} However, a fit of good quality can be made to both $Z$ boson and
low-energy Drell-Yan data once a small number of free parameters is
allowed in the fit to describe the unknown power corrections \cite{Landry:2002ix}.
Variations of these parameters lead to small changes in the $q_{T}$
spectrum of $Z$ bosons and large changes in the low-$Q$ Drell-Yan
distributions. The $Z$ boson data taken alone also does not constrain
the dependence of the nonperturbative terms on $Q$ and parton flavor.
Consequently a combination of Drell-Yan and $Z$ boson data imposes
tighter constraints on the nonperturbative parameters than the $Z$
boson data by itself, pretty much as a combination of the inclusive
data from various experiments places better constraints on the PDF's
than each individual experiment.

The feasibility of such a global $p_{T}$ fit crucially relies on
the universality of the nonperturbative contributions. Whether the
universality holds is a question under active investigation. Universality
of the large-$b$ contributions follows from the factorization theorem
for the resummed cross sections, which was stated in \cite{Collins:1985kg}
and recently proved in \cite{Collins:2004nx}; see also the related
discussion in Ref.~\cite{Ji:2004xq}. A high-quality fit to $p_{T}$
data from diverse experiments would provide, in principle, empirical
confirmation of the universality. The fits utilizing the $b_{*}$
anzatz \cite{Collins:1985kg} -- the simplest model of the nonperturbative
terms -- can indeed achieve a decent $\chi^{2}$ of $170$ ($130$)
per 120 data points for the parameter $b_{max}$ equal to 0.5 (0.8)
$\mbox{{ GeV}}^{-1}$. However, the issue needs further investigation
because of the substantial uncertainties in the interpretation of
the fits, mostly associated with the low-energy Drell-Yan data. The
preferred form of the nonperturbative function is strongly correlated
with the overall normalizations of the Drell-Yan cross sections, which,
in their turn, are affected by ${\cal O}(\alpha_{s}^{2})$ corrections
(of order $10-15\%$ \cite{Anastasiou:2003yy}), the PDF's, and experimental
uncertainties. The true uncertainty in the nonperturbative function
is linked to the rest of the factors, which are not under sufficient
control yet in the Drell-Yan process. This is particularly true with
respect to the PDF parameters, which in principle should be fitted
together with the nonperturbative function to obviate correlations
between the shape of $d\sigma/dq_{T}$ and parton densities. 

\begin{figure}
\includegraphics[%
  height=6cm,
  keepaspectratio]{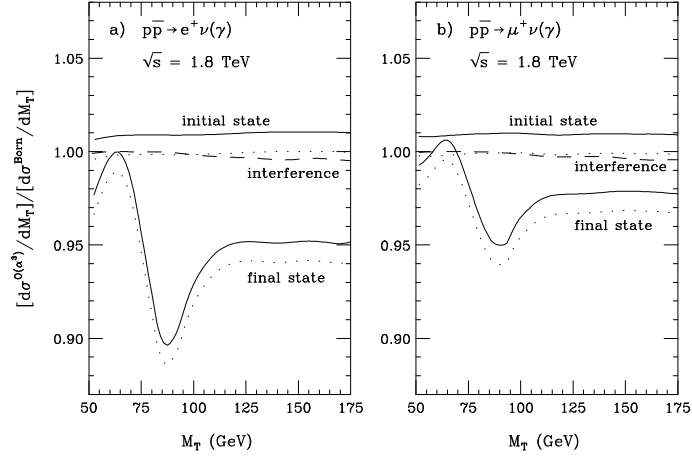}

\caption{\label{fig:EWMT} The ${\cal O}(\alpha)$ electroweak corrections
to the leptonic transverse mass distribution, $d\sigma/dM_{T}^{\ell\nu}$,
in $W$ boson production \cite{Baur:1998kt}. No leptonic detection
constraints are imposed.}
\end{figure}

\textbf{Electroweak corrections to $M_{W}$} are dominated by the
QED radiation from the final-state charged lepton, which results in
some loss of the charged lepton's momentum to the surrounding cloud
of soft and collinear photons. The final-state QED (FQED) radiation
changes the extracted value of $M_{W}$ by shifting the Jacobian peak
in the $M_{T}^{\ell\nu}$ distribution in the negative direction {[}cf.~Fig.~\ref{fig:EWMT}{]}.
In contrast, the initial-state radiation and interference terms mostly
change the overall normalization of the Jacobian peak and have a smaller
effect on $M_{W}$. The size of the electroweak corrections depends
on the mode of lepton identification. The collinear photons are merged
with the electrons in the calorimeter towers, but isolated from the
muons in the muon detectors. Consequently the corrections in the electron
channel are larger than in the muon channel in the absence of lepton
identification requirements, but smaller once the lepton identification
is taken into the account. The CDF Run-1b analysis estimates the shift
in the $W$ boson mass due to the FQED radiation by $-65\pm20$ MeV
and $-168\pm10$ MeV in the $e\nu$ and $\mu\nu$ decay channels,
respectively \cite{Affolder:2000bp}. The errors of 20 and 10 MeV
correspond to the radiative corrections that were ignored in the Run-1
analysis, such as the ${\cal {\cal O}}(\alpha)$ initial-state radiation
and interference, or radiation of two and more photons. These errors
need to be reduced to push the total $\delta M_{W}$ down to 30 MeV
in the Run-2 measurement.

\textbf{The correlations between $q_{T}$ spectrum, partonic structure,
and electroweak radiation} may prove to be consequential for the upcoming
analyses, given that all three are the major sources of uncertainties
in the extracted $W$ boson mass. In Run-1, these interconnections
were mostly discarded, so that, for instance, simulations of the QCD
radiation and electroweak radiation were performed by independent
computer programs \textsc{(ResBos} \cite{Balazs:1997xd,Landry:2002ix}
and \textsc{W/ZGrad} \cite{Baur:2001ze,Baur:1998kt}) and combined
\emph{ad hoc} in the final output. 

Starting from the Run-2, the correlations between the dominant factors
cannot be dismissed. The FQED radiation contributes the bulk of the
full electroweak correction, and it cross-talks with the initial-state
QCD radiation through conservation of momentum and spin. The combined
effect of the ${\cal O}(\alpha)$ FQED correction and the resummed
QCD correction was estimated for the Run-2 observables by using a
new computer program \textsc{ResBos-A} (\textsc{ResBos} with FQED
effects) \cite{Cao:2004yy}. The FQED and resummed QCD corrections
to the Born-level shape of the Jacobian peak in the $M_{T}^{\ell\nu}$
distribution were found to be approximately (but not completely) independent.
The reason is that the $M_{T}^{\ell\nu}$ distribution is almost invariant
with respect to the transverse momentum of $W$ bosons, so that the
QCD correction reduces, to the first approximation, to rescaling of
the Born-level $M_{T}^{\ell\nu}$ distribution by a constant factor.
The relationship between FQED and QCD corrections is more involved
in the leptonic $p_{T}$ distributions, which depend linearly on $q_{T}$
of $W$ bosons. In the $p_{T\ell}$ channel, the combined effect does
not factorize into separate FQED and QCD corrections to the Born-level
cross section.  Additional modifications will be caused by the FQED
correction to the finite part of the resummed cross section ($Y$-piece)
and finite resolution of the detector, which were not considered in
Ref.~\cite{Cao:2004yy}.

\begin{figure}
\includegraphics[%
  height=7cm,
  keepaspectratio]{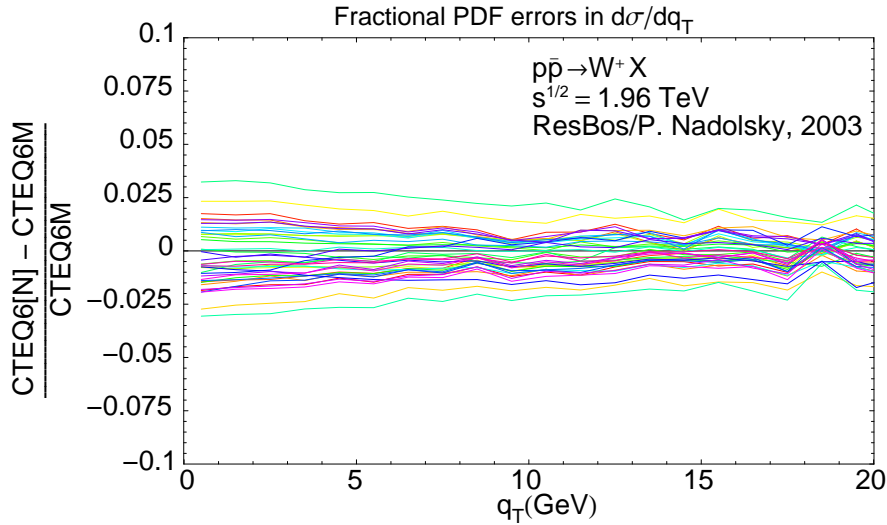}

\caption{\label{fig:qTPDF} The PDF uncertainty in the resummed $q_{T}$ distribution
for $W$ boson production in the Tevatron Run-2, evaluated using CTEQ6
PDF's \cite{Pumplin:2002vw} and BLNY form of $S_{NP}(b)$ \cite{Landry:2002ix}.
The lines show the fractional differences between $d\sigma/dq_{T}$
for 40 sample PDF sets CTEQ6{[}N{]} ($N=1,\,40$), and the central
PDF set (CTEQ6M).}
\end{figure}

The PDF uncertainties were estimated in Run-1 by repeating the analysis
for select sets of parton densities, which did not cover the full
span of allowed variations in the PDF parameters. A more systematical
estimate can be realized by applying the new techniques for the PDF
error analysis. The choice of the PDF set affects $q_{T}$distributions
directly, by changing the PDF's in the factorized cross section, but
also indirectly, by modifying the nonperturbative Sudakov function
$S_{NP}(b)$ in the resummed form factor. For a chosen form of $S_{NP}(b)$,
the PDF errors can be evaluated within the Hessian matrix method,
by repeating the computation of $q_{T}$ distributions for an ensemble
of sample PDF sets. The variations in the resummed $q_{T}$ spectrum
for 41 CTEQ6 PDF sets \cite{Pumplin:2002vw} and BLNY nonperturbative
Sudakov function \cite{Landry:2002ix} are shown in Fig.~\ref{fig:qTPDF}.
Depending on the choice of the PDF set, $d\sigma/dq_{T}$ changes
by up to $\pm3\%$ from its value for the central PDF set (CTEQ6M).
The variations in the PDF's modify \emph{both} the normalization and
shape of $d\sigma/dq_{T}$. Although the changes in the shape are
relatively weak at $q_{T}<10$ GeV, they cannot be ignored when $M_{W}$
is extracted from the $p_{T\ell}$ distribution. These results do
not reflect possible correlations between the PDF's and $S_{NP}(b)$
in the global fit to $p_{T}$ data, which may be introduced by the
dependence of $S_{NP}(b)$ on the normalizations of the low-$Q$ Drell-Yan
cross sections. In contrast to the PDF's, $S_{NP}(b)$ does appreciably
modify the shape of $d\sigma/dq_{T}$ and may cause larger shifts
in the extracted $M_{W}$. As discussed earlier, the correlation between
free parameters in the PDF's and $S_{NP}(b)$ could be explored by
performing a simultaneous global analysis of the inclusive cross sections
and $p_{T}$-dependent data. The first steps towards realization of
such a combined fit are being taken within the CTEQ collaboration
\cite{HustonSigTot}. 

The theoretical assumptions need further scrutiny as well, to guarantee
their up-to-date level and explore their validity in the new phase
space regions accessible in the Run-2 and at the LHC. For example,
substantial deviations from the present model may occur in the resummed
cross sections at momentum fractions $x$ of order $10^{-2}$ or less,
typical for forward rapidities in the Run-2, and for all rapidities
at the LHC \cite{Berge:2004nt}. Such deviations may be caused by
enhancement of $\ln(1/x)$ terms in the perturbative or nonperturbative
parts of the resummed form factor, which could lead to the broadening
of $q_{T}$ distributions at small $x$. An estimate based on the
analysis of small-$x$ semi-inclusive hadroproduction at HERA suggests
that the $q_{T}$ broadening may be visible in a sample of forward-rapidity
$Z$ bosons collected in the Run-2. If found at the Tevatron, the
$q_{T}$ broadening will have a profound impact on $W$ and $Z$ production
at the LHC. 

Distinctions between the quark flavors were also neglected in the
previous resummation studies, with the exception of the explicit flavor
dependence in the parton densities. This assumption is likely violated
at some level, especially for the charm and bottom quarks, whose heavy
masses suppress very soft radiation. While the heavy-quark subprocesses
are rare at the Tevatron, they become important at the LHC, where
about $17\%$ ($24\%$) of $W^{+}$($W^{-}$) bosons will be produced
in scattering of charm quarks. The heavy-quark masses can be incorporated
into the resummation in an extension of the CSS formalism to the massive
variable-flavor number factorization scheme \cite{Nadolsky:2002jr}.
We estimate within this extension that the heavy-flavor mass terms
will have essentially no effect on the extracted $M_{W}$ in the Tevatron
Run-2, but may shift $M_{W}$ by $\sim5$ MeV (depending on the size
of the heavy-flavor nonperturbative terms) at the LHC \cite{BergeHeavyQuark}.

\textbf{Conclusions.} The theoretical model for $W$ and $Z$ boson
production at hadron colliders undergoes rapid development. The ${\cal O}(\alpha_{s}^{2})$
QCD corrections are becoming available for many $W$ and $Z$ observables.
The ${\cal O}(\alpha_{s}^{2})$ corrections show tiny scale dependence
and can be approximated in several important cases by a uniform rescaling
of the ${\cal O}(\alpha_{s})$ cross section by $2.5-5\%$. The ${\cal O}(\alpha_{s}^{2})$
correction cannot be replaced by a constant $K$-factor when the lowest-order
contribution enters at ${\cal O}(\alpha_{s})$ (e.g., at large $q_{T}$
or in leptonic angular distributions), or near the edges of phase
space (e.g., at forward rapidities). The finite-order cross sections
have to be improved by resummation in kinematical limits with enhanced
higher-order radiation, notably at $q_{T}\ll Q$, but also at $q_{T}\gg Q$,
$Q\gg M_{V}$, $x\rightarrow0$, or $x\rightarrow1$. Both formal
and phenomenological aspects of the resummation formalisms require
further investigation.

The ${\cal O}(\alpha)$ electroweak corrections are also available,
and they should be applied together with the QCD corrections to reach
the required accuracy. Correlations between various dynamical factors
(PDF's, $q_{T}$ power corrections, electroweak radiation, etc.) may
be important in the full context and must be systematically explored.
Finally, the newly found analytical results must be optimally implemented
in numerical simulations, which should be both fast and precise to
adequately serve various practical uses. 


\begin{theacknowledgments}

I thank my colleagues at Argonne National Laboratory, Michigan State University, and Southern Methodist University for collaboration and illuminating discussions. This work was supported in part by the US Department of Energy, High Energy Physics Division, under Contract W-31-109-ENG-38.

\end{theacknowledgments}


\end{document}